\documentstyle[pre,aps,epsf]{revtex}

\begin{document}
\draft

%
\title{Interaction-driven equilibrium and statistical laws 
in small systems. The cerium atom.} 


\author{V. V. Flambaum, A. A. Gribakina, 
        G. F. Gribakin and I. V. Ponomarev.}
 

\address{ School of Physics, The University of New South Wales,
Sydney 2052, Australia}

\date{\today}
\maketitle

\begin{abstract}
It is shown that statistical mechanics is applicable to isolated quantum
systems with finite numbers of particles, such as complex atoms, atomic
clusters, or quantum dots in solids, where the residual two-body
interaction is sufficiently
strong. This interaction mixes the unperturbed shell-model (Hartree-Fock)
basis states and produces chaotic many-body eigenstates. As a result,
an interaction-induced statistical equilibrium emerges in the system.
This equilibrium is due to the off-diagonal matrix elements of the
Hamiltonian. We show that it can be described by means of temperature
introduced through the canonical-type distribution. However,
the interaction between the particles can lead to prominent deviations
of the equilibrium distribution of the occupation numbers from the
Fermi-Dirac shape. Besides that, the off-diagonal part of the Hamiltonian
gives rise to the increase of the effective temperature of the system
(statistical effect of the interaction). For
example, this takes place in the cerium atom which has four valence
electrons and which is used in our work to compare the theory with
realistic numerical calculations.
\end{abstract}

\section{Introduction}\label{sec:1}
Consider a many-body quantum system of interacting particles.
If the number of particles is large, statistical laws can be applied to
describe the properties of the system. They can also be applied to
few-particle systems (or even single particles) interacting with a heat
bath.
In both cases the equilibrium is achieved at arbitrarily weak interaction
between the particles, or with the heat bath.
If the interaction between the particles is strong enough, a different
kind of statistical equilibrium is possible in isolated few-particle
quantum systems. It is achieved when the excited states of the system
become chaotic {\em compound} states. The systems examined so far are the
rare-earth atom of Ce \cite{Ce} with just
four active valence electrons, 12 nucleons in the $s-d$ shell \cite{Zel},
and $n=4$--7 particles interacting by means of a two-body random
interaction \cite{FIC96,FI97}. In spite of the obvious differences these
systems have much in common, as far as properties of their eigenstates
and various statistics are concerned. It has been shown in
\cite{Zel,FIC96,FI97} that in the regime
of compound excited states one can introduce such thermodynamic parameters
as temperature and entropy, and observe other typically statistical
features, e.g., the Fermi-Dirac distribution of the occupation numbers.

In the present work we concentrate on the
statistics of the occupation numbers in a realistic Fermi system: the
atom of Ce. We show that when the interaction between the particles is
strong and the two-body matrix element fluctuates strongly as function
of the single-particle states involved (hence, there is no good mean-field
approximation), large deviations from the
Fermi-Dirac behaviour are observed. However, a statistical description of
the system including the introduction of a temperature is still possible
if the interaction between the particles is properly accounted for.

The notion of compound states is important for our work, so we would like
to explain it in greater detail. Suppose that for a given range of
excitation energies the interaction between the particles gives rise to a
certain mean field. This mean field can then be used to generate a set of
single-particle orbitals. The multiparticle basis states of the system can
be constructed from these orbitals by simply specifying their
occupation numbers. The spectrum of such states in a system with
several active particles is very dense since there are many ways of 
distributing them among the orbitals. For the interacting particles
these multiparticle states are not the eigenstates of the system. Instead,
they are mixed together by the residual two-body
interaction. If this interaction is strong, the number of basis states
``involved'' in almost every eigenstate of the system becomes very large
(about 100 in atoms and up to $10^6$ in nuclei), and the eigenstates become
almost random (chaotic) superpositions of the basis states, devoid of any
good quantum numbers, save the exact ones -- energy, parity,
total angular momentum, etc. Following the nuclear physics terminology
we call such eigenstates {\em compound states}.
Their statistical properties, e.g., distribution over the unperturbed
basis states, are very similar in different systems studied: atoms,
nuclei, or a two-body random interaction model. Most importantly,
they provide a good starting point for developing a statistical theory
for isolated few-particle systems \cite{FI97}.

\section{Statistics of the occupation numbers}\label{sec:2}

Statistical behaviour is usually established in the limit of
a large number of particles $n$. Moreover, simple quantitative results can
be obtained if correlations between the particles are somehow weak.
This means 
that the interaction between the particles can be neglected, or --
more realistically -- an appropriate mean field theory is developed.
The latter results in the picture of free quasiparticles moving in the
effective self-consistent field created by the constituents.

In the limit of large $n$ the temperature $T$ is a well-defined physical 
quantity and all equilibrium characteristics can be found by applying
the canonical (Gibbs) probability distribution $w_i\propto \exp (-E^{(i)}/T)$,
where $E^{(i)}$ is the energy of the $i$th eigenstate of the system.
For example, for a gas of noninteracting fermions this results in
the famous Fermi-Dirac distribution (FDD) of the occupation numbers:
\begin{equation}\label{FDd}
\overline{n}_\alpha =\frac{1}{\exp [(\varepsilon _\alpha -\mu)/T]+1}~,
\end{equation}
where $\varepsilon _\alpha $ is the energy of a single-particle state
$\alpha $, and $\mu=\mu(n,T)$ is the chemical potential. It is
determined, at a given temperature, from the normalization condition
$\sum_\alpha \overline{n}_\alpha = n$. Formula (\ref{FDd}), when it is
valid, in fact provides one with a relation between the temperature and
the energy $E$ of the system:
$\sum_\alpha \varepsilon _\alpha \overline{n}_\alpha =E$, which can
also be viewed as a possible definition of the temperature. Note that
Eq. (\ref{FDd}) can hold for the interacting fermions as well, provided
we consider the distribution of {\em quasiparticles}, and replace
$\varepsilon _\alpha $ with the quasiparticle energy
$\tilde \varepsilon _\alpha $, which in turn depends on the distribution
of excited quasiparticles. This is an important point of Landau's theory
of Fermi liquids (see e.g., \cite{PiNo}). 

Strictly speaking, the FDD is derived for the grand canonical ensemble,
where $\mu $ is fixed, and $n$ is the mean number of particles \cite{LL5}.
For large $n$ the difference is negligible unless fluctuations of the
number of particles are concerned. In appendix \ref{app} we consider the
approximations one has to make to arrive at the FDD (\ref{FDd}) for a
finite system of $n$ interacting particles obeying the canonical
distribution.

In reality there are many complex systems, such as compound nuclei,
rare-earth atoms, molecules, atomic clusters, or quantum dots, which do not
satisfy the conditions for Eq. (\ref{FDd}) to hold. However, their
complexity suggests that some statistical methods can be developed, and in
nuclear physics such statistical temperature-based description has been
known for quite a while. Intuitively such description is very natural, and
a more rigorous justification does not seem to have been needed.

The number of active particles in these systems can be relatively small
($\lesssim 10$), whereas the interaction between them (even the residual
interaction in the mean-field basis) is large, i.e., greater than the
energy intervals between unperturbed many-particle basis states.
This interaction makes up for the absence of a heat bath, and promotes
the onset of ``randomization'' and quantum chaos. This chaos is purely
dynamical, in the sense that the Hamiltonian matrix of the system
does not contain any random parameters, yet, the behaviour can be
complicated enough (``chaotic''), and a number of properties, e.g.,
the energy level statistics, are consistent with the predictions of
random matrix theories \cite{Ce,Zel,FIC96}.
This gives one a possibility to talk about some kind of equilibrium 
in the system, and pursue the development of a statistical theory for
few-body Fermi systems \cite{FI97}.

In what follows we will look at the results obtained numerically in
a realistic model of the Ce atom which contains only four
active particles (valence electrons). We will see that the energy
dependence of the occupation numbers differs prominently from what one
expects from the FDD (\ref{FDd}), and show that this behaviour results
from the strong fluctuations of the two-body Coulomb interaction for
different orbitals. We then show that this interaction
can be taken into account within the statistical approach to calculation
of the occupation numbers and other mean values, leading to a good
agreement between the results of our {\em statistical} theory and the
numerical calculations. Such agreement confirms the existence of
equilibrium similar to the thermal one in the system of a few strongly
interacting particles.

\section{The Ce atom}\label{sec:3}

The cerium atom has one of the most complicated spectra in the periodic
table. The density of energy levels with a given total angular momentum and
parity $J^\pi $ reaches hundreds of levels per eV at excitation
energies of just a few eV, well below the ionization threshold of
$I=5.539$ eV
\cite{tables,AIS}. The Ce atom has four valence electrons, and a well
defined
$4f6s^25d$ ground state. However, with the increase of the excitation
energy and involvement of yet another low-lying electron orbital $6p$,
the atomic eigenstates become compound states (in the sense of Sec.
\ref{sec:1}), and it becomes absolutely impossible to choose any reasonable
coupling scheme or provide any classification for them. At the same time,
the orbital occupation numbers move away from integer values, and even the
idea of a dominant configuration for a particular energy level becomes
meaningless \cite{Ce}.

In the present work we continue to study the cerium atom numerically, with
emphasis on the energy dependence of the occupation numbers.
The electronic structure of the Ce atom consists of a Xe-like
$1s^2\dots 5p^6$ core and four valence electrons. A large difference in the
energy scales of the core and valence electrons allows us to neglect
excitations from the core and consider the wave function of the core as a
``vacuum'' state $|0\rangle$. Accordingly, the four active
electrons added to this vacuum form the spectrum of Ce at excitation
energies $E\lesssim I$ \cite{corex}.

The calculations are performed using the Hartree-Fock-Dirac (HFD) and 
configuration interaction (CI) methods (see \cite{Ce} for details).
A self-consistent HFD calculation of the neutral atom results in the
construction of the core and valence orbitals. It also determines the
mean-field potential, which is then used to calculate the basis set
of single-particle ortho-normalized relativistic states
$|\alpha \rangle = |nljj_z\rangle $ with energies $\varepsilon _\alpha$.
This procedure defines the zeroth-order Hamiltonian of the system,
\begin{equation}\label{H0}
\hat{H}^{(0)}=\sum_{\alpha}\varepsilon_\alpha a^\dagger _\alpha
a_\alpha ~.
\end{equation}

The unperturbed multi-particle basis states $|k\rangle $ constructed
from the single-particle states, $|k\rangle = a^\dagger _{\nu_1}
a^\dagger _{\nu_2}a^\dagger _{\nu_3}a^\dagger _{\nu_4}|0\rangle $,
are eigenstates of $\hat H^{(0)}$:
$\hat{H}^{(0)}|k \rangle = E^{(0)}_k|k\rangle $, where
\begin{equation}\label{E0}
E^{(0)}_k = \sum _\alpha \varepsilon_{\alpha}n^{(k)}_{\alpha}
\end{equation}
is the zeroth-order energy of the state $|k\rangle $,
and  $n^{(k)}_{\alpha}=\langle k|\hat{n}_\alpha |k\rangle =
\langle k|a^\dagger _\alpha a_\alpha |k\rangle $ are the occupation
numbers equal to 0 or 1, depending on whether the state $\alpha $
is occupied in $|k \rangle $, or not. To subtract additional symmetries
only the basis states $|k\rangle $ with a given projection of the total
angular momentum $J_z$ and parity are considered.

The total Hamiltonian $\hat H$ of the active electrons is the sum of the
zeroth-order mean-field Hamiltonian $\hat H^{(0)}$ and the 2-body residual
interaction
\begin{equation}\label{V}
\widehat{V}=\frac{1}{2}\sum _{\alpha \beta \gamma \delta }
V_{\alpha \beta \gamma \delta }
a^\dagger _\alpha a^\dagger _\beta a_\gamma a_\delta ~.
\end{equation}
The residual interaction $\widehat{V}$ contributes to the diagonal and
off-diagonal matrix elements between the multiparticle states $|k\rangle $.
Its diagonal part shifts the energy of the basis state $k$,
\begin{equation}\label{Vk}
\Delta E_k=V_{kk}=\sum_{\alpha >\beta}\left( V_{\alpha \beta \beta \alpha }
-V_{\alpha \beta \alpha \beta }\right) n^{(k)}_\alpha n^{(k)}_\beta ~.
\end{equation}
The off-diagonal matrix elements $V_{k'k}=
\langle k'|\widehat V |k\rangle $ are responsible for mixing of the
multiparticle basis states.

Complete diagonalization of the operator $\hat{H}=\hat{H_d}+\widehat{V}$
in the space of the basis states  $|k \rangle $ yields ``exact''
energies $E^{(i)}$ and stationary states $|i\rangle $,
\begin{equation}\label{fulldia}
\hat{H} |i\rangle ={\em E^{(i)}} |i\rangle ~,
\end{equation}
which can be presented as superpositions of the unperturbed basis states,
\begin{equation}\label{expan}
|i\rangle =\sum_k C_k^{(i)}|k\rangle ~,\quad \sum_k
\left|C_k^{(i)}\right| ^2=1~.
\end{equation}

In this work we included 14 relativistic subshells $nlj$ in the
single-particle basis ($6s$, $7s$, $6p$, $7p$, $5d$, $6d$, $4f$, and
$5f$), and performed exact diagonalization of the
$N\!\times \!N$ Hamiltonian matrix in a Hilbert space with
$N\sim 8\times 10^3$, obtained by truncating the complete set of the
shell-model atomic configurations. Our numerical
results are obtained for the even states of Ce with the total angular
momentum projection set to $J_z=0$. Thereby, all possible states with $J$
from 0 to 10 are included. For the given choice of the basis the numbers
of eigenstates with $J=0$--10 are 343, 917, 1354, 1493, 1433, 1153, 826,
497, 262, 107, and 34, respectively.

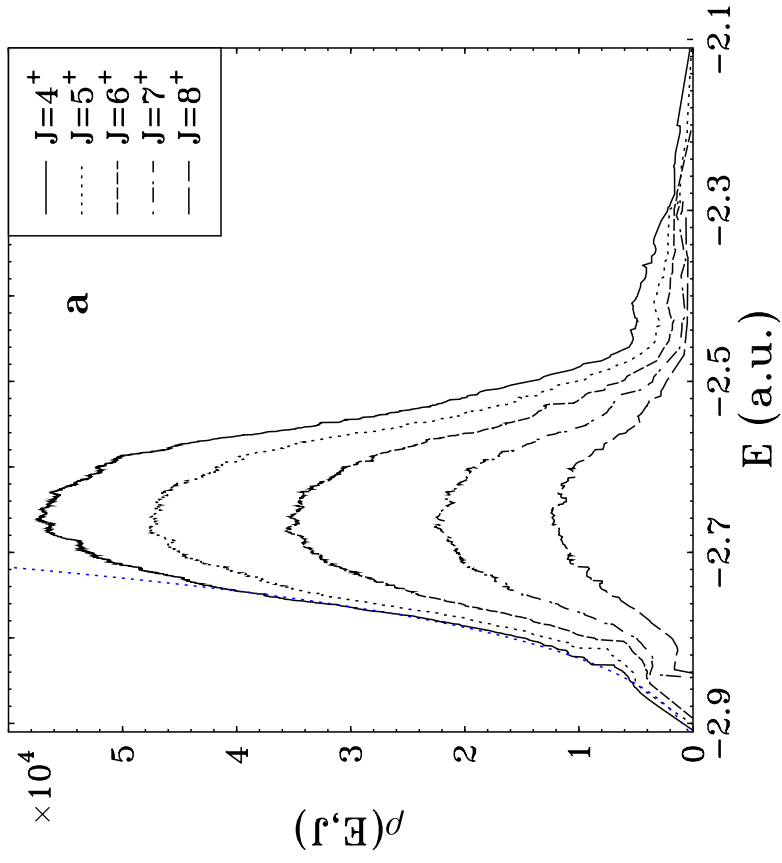
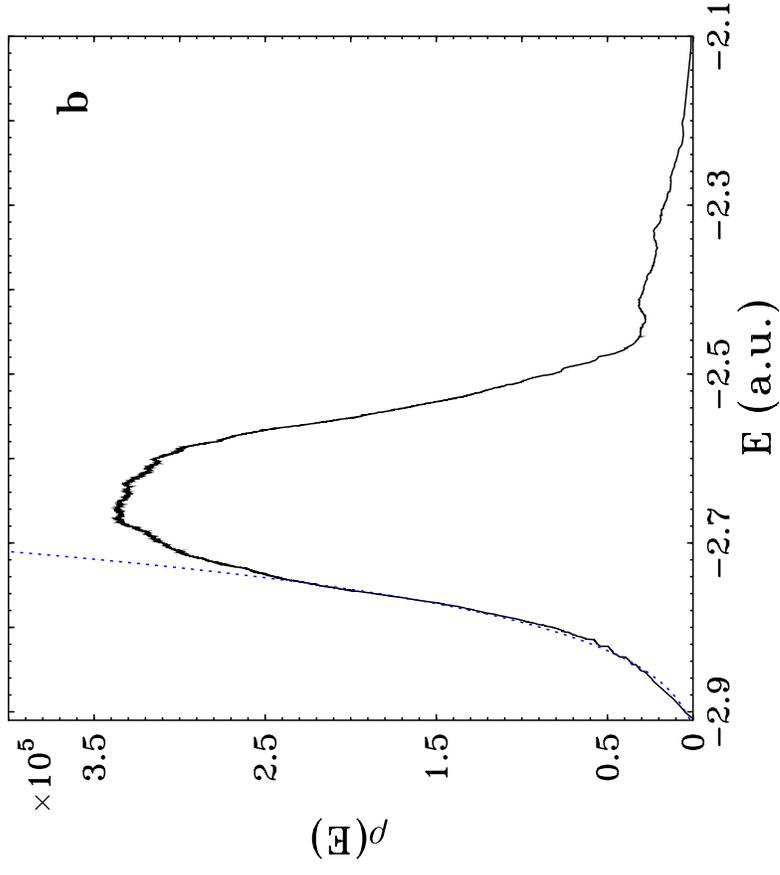
\begin{figure}
\caption{Eigenvalue densities for the even states of Ce averaged over the
energy interval $\Delta E=0.05$ a.u.
a. The upper solid curve is for $J=4$, and lower curves correspond
to successively increasing values of $J$. Note that all densities have
similar shapes. Dotted curve is the analytical fit for $J=4$:
$\rho (E)=\rho _0\exp \left[ a(E-E_g)^{1/2}\right]$, where
$\rho _0=119$, $a=12.3$ a.u. and $E_g=-2.91$ a.u.
b. Total level density of the even states and fit with
$\rho _0=27$, $a=13.0$ a.u., and $E_g=-2.91$ a.u. \label{J49}}
\end{figure}

The eigenvalue densities $\rho _J (E)$ for $J=4$--8 are shown in
Fig. \ref{J49}. They have been window-averaged over
$\Delta E=0.05$ a.u. to smooth out short-range fluctuations. The largest
density is observed for $J=4$, and with the exception of small regions
near the ends of the spectra, all $\rho _J (E)$ are proportional to each
other. The shapes of the eigenvalue densities are basically determined by
the corresponding basis state densities (although the effect of level
repulsion
makes the former slightly wider). They are characterized by a very rapid
increase in the low-energy part. This increase is a direct consequence of
the fact that the accessible energy can be distributed in an ever greater
number of ways between the 4 electrons. Being essentially
of combinatorial nature, the level density can be described by the exponent
\begin{equation}\label{rhoE}
\rho (E)\propto \exp \left( a\sqrt{E-E_g}\right)
\end{equation}
which is derived in the independent-particle model
\cite{BM} ($E_g$ is the ground state energy). As seen from Fig. \ref{J49}
the level density from our calculation indeed follows (\ref{rhoE}) at
low energies, but then reaches its maximum and decreases \cite{French}.
These latter
features are unphysical as they are consequences of the finite size
of our basis. However, within 5 eV of the ground state the configuration
set we use is reasonably complete.

Relativistic atomic subshells $nlj$ are $(2j\!+\!1)$-degenerate,
therefore, we consider average occupation numbers
\begin{equation}\label{ns}
\hat n_s=g_s^{-1}\sum _{\alpha \i functionn s}\hat n_\alpha =
g_s^{-1}\sum _{\alpha \in s}a^\dagger _\alpha a_\alpha ~,
\end{equation}
where $g_s=2j+1$ is the degeneracy of the subshell $s$. In the
system of a large number of weakly interacting particles thermally
averaged values of $n_s$ are given by the FDD (\ref{FDd}). In the quantum
dynamical system, like the Ce atom, the occupation numbers for any
eigenstate can be
obtained as $n_s^{(i)}=\langle i|\hat n_s|i\rangle $. When the number
of active particles is small $n_s^{(i)}$ show strong level-to-level
fluctuations, and it is more instructive to look at the spectrally
averaged values
\begin{equation}\label{nsE}
n_s(E)=\overline{\langle i|\hat n_s|i\rangle }=
\sum_k \overline{\left| C_k^{(i)}\right| ^2}\langle k|\hat n_s|k \rangle ~,
\end{equation}
where overline means averaging over the eigenstates $i$ within
some energy interval around $E$.

A typical distribution of the occupation numbers calculated at the
excitation energy of 3.75 eV above the atomic ground state is shown in
Fig.~\ref{nfdsp} as a function of the single-particle energy
$\varepsilon _s$ of the orbitals (see Table \ref{param}). The values of
$\varepsilon _s$ are determined with respect to the Xe-like Ce$^{4+}$
core. One can see that the distribution does
not look at all like a monotonically decreasing FDD.
A similar picture is observed over the whole energy interval from
the ground state to the ionization potential. For example, the lowest
even state of Ce has a configuration of $4f^26s^2$, while the FDD would
tell us that all 4 electrons must be placed in the lowest $4f$-orbital,
when the energy or ``temperature'' of the system is low. In reality the
$4f^4$ electron configuration lies at very high energies due to a strong
electron repulsion in this compact orbital (the radius of the $4f$ orbital
in Ce is at least two times smaller than that of any other valence
orbital).

Of course, considering the orbital energies $\varepsilon _s$ has the
drawback that they completely ignore the residual interaction between the
valence
electrons. When this interaction is strong one would wish to introduce
some new mean-field orbital energies $\tilde \varepsilon _s$ that would
incorporate the effect of such interaction. The value of
$\tilde \varepsilon _s$ for the orbital $s$ will inevitably depend on
the distribution of the other electrons, and hence on the excitation
energy of the system. In Sec. \ref{sec:5} we introduce such energies
within the statistical approach. However, even when the occupation numbers
are plotted against $\tilde \varepsilon _s$ there is still a large
deviation from the standard FDD distribution.

At first sight such a strong deviation from the FDD in a strongly
interacting Fermi system speaks against any possibility of a statistical
description of the system. In what follows we show that the strong Coulomb
interaction between the electrons can be incorporated in the
canonical-ensemble description of the system, and thermally averaged
occupation numbers can be $n_s(T)$ derived. As a result of the electron
interaction, the distribution of $n_s(T)$ differs from the FDD. What is
more important, we find good agreement between the results obtained by
means of this statistical approach and those from the pure dynamic
calculation of the Ce eigenstates, $n_s(T)\approx n_s(E)$. We also find a
way to relate the energy and the temperature.

\section{Canonical ensemble and the strength function}\label{sec:4}

In this section we show that averaging over the {\em canonical
distribution},
which weighs different states according to their energies $E_k$ with
probabilities $w_k\propto \exp (-E_k/T)$, is very similar to averaging
over the
exact eigenstates $|i\rangle $, when these eigenstates are {\em compound},
i.e., include large numbers $N_c$ of basis states $|k\rangle $ mixed
together with small weights $C_k^{(i)}\sim 1/\sqrt{N_c}$,
Eq.~(\ref{expan}). For a classical system the latter is equivalent to
averaging over the {\em microcanonical distribution} that considers
all points on the surface $E\!=\!\mbox{const}$ of the phase space as
equally probable \cite{FV93}. As is known, the two types of averaging yield
identical results for large systems \cite{LL5}.

\subsection{Averaging over the canonical distribution at a given
temperature.}  
\label{avcan}

Suppose first that the off-diagonal part of the residual two-body
interaction $\widehat V$ is switched off. The multiparticle basis states
$|k\rangle $ then correspond to the stationary states of the system
with energies $E_k=E_k^{(0)}+\Delta E_k$. The interaction with a heat
bath at temperature $T$ results in the canonical distribution of
probabilities of finding the system in a given state $k$,
$w_k=Z^{-1}\exp (-E_k/T)$, where $Z=\sum _k \exp (-E_k/T)$, so that
$\sum _kw_k=1$. The occupation numbers at
temperature $T$ are calculated as
\begin{equation}\label{nsT}
n_s(T)=\sum _kw_k\langle k|\hat n_s|k\rangle =
Z^{-1}\sum _k\exp (-E_k/T) \langle k|\hat n_s|k\rangle ~.
\end{equation}

The spectrum of $E_k$ is similar to the eigenvalue spectra shown in
Fig.~\ref{J49}, and is characterized by a rapid rise of its density
[see Eq. (\ref{rhoE})]. Thus, if we replace summation in
Eq. (\ref{nsT}) by integration over $E_k$,
\begin{equation}\label{nsTint}
n_s(T)=\int w_T(E_k)
\langle k|\hat n_s|k\rangle dE_k ~,
\end{equation}
where we introduced the probability density
$w_T(E_k)=Z^{-1}\exp (-E_k/T)\rho (E_k)$, the integrand in (\ref{nsTint})
will peak strongly due to competition between the two exponents, the
decreasing $\exp (-E_k/T)$ and the rising $\rho (E_k)$. As a result, the
main contribution to $n_s(T)$ is given by the vicinity of the maximum of
$w_T(E_k)$. The equation for the position of the maximum 
$E_k=E$ provides a relation between the most probable energy $E$ and the
temperature,
\begin{equation}\label{defT}
-\frac{1}{T}+\frac{d\{ \ln [\rho (E)]\}}{dE}=0~.
\end{equation}
If the temperature is not too small the maximum of $w_T(E_k)$ is almost
symmetric, and the most probable energy becomes close to the mean energy:
\begin{equation}\label{meanE}
E\approx \overline{E}(T)=\int E_kw_T(E_k)dE_k ~.
\end{equation}
If we use the analytical form (\ref{rhoE}), Eq. (\ref{defT}) yields
\begin{equation}\label{TE}
T=\frac{2}{a}\sqrt{E-E_g}~.
\end{equation}

\subsection{Strength function and averaging over the compound states.}

Let us now come back to the dynamic description of the isolated many-body
quantum system and switch on the off-diagonal part of the residual
interaction. In this case the eigenstates are given by Eq. (\ref{fulldia})
and the occupation numbers at a given energy are found from
Eq. (\ref{nsE}). The key quantity in calculating $n_s(E)$ is the
mean-squared eigenstate component $\overline{\left| C_k^{(i)}\right| ^2}$.
When $\widehat V$ is strong enough and the energy $E^{(i)}$ is not too close
to the ground state, $\overline{\left| C_k^{(i)}\right| ^2}$ represents
the spreading of the eigenstate over a large number of basis states. It is
proportional to the strength function (introduced by Wigner \cite{wigner}
and also known as the local density of states),
\begin{equation}\label{LDOS}
\rho _w(E,k)=\sum _i\left| C_k^{(i)}\right| ^2 \delta (E-E^{(i)})
\simeq \overline{\left| C_k^{(i)}\right| ^2}\rho (E)~,
\end{equation}
where $\rho (E)$ is the eigenvalue density. The last equality in
(\ref{LDOS}) implies that some averaging over the the energy interval
greater than the level spacing has been performed at $E^{(i)}\approx E$.
It follows
from numerical calculations \cite{Ce,Zel,FI97} as well as from analytical 
considerations \cite{BM,wigner,CFIC} that $\rho _w(E,k)$ is a bell-shaped
function
centered at $E\approx E_k$. Near its maximum it depends only on the
difference $E-E_k$, and can be described by the Breit-Wigner formula
\begin{equation}\label{BW}
\rho _w(E,k)=\frac{\Gamma /2\pi }{(E-E_k)^2+\Gamma ^2/4}~.
\end{equation}
The {\em spreading width} $\Gamma $ is usually given by
$\Gamma \simeq 2\pi \overline{|\langle k'|\widehat V|k\rangle |^2}
\rho (E)$. Therefore, the basis states are strongly mixed together by the
residual interaction only locally, within the energy range $\Gamma $.

The notions of the strength function and the spreading width become
meaningful if the interaction is strong enough, and $\Gamma \gg D$,
where $D=1/\rho (E)$ is the mean level spacing (accordingly,
$\overline{|\langle k'|V|k\rangle |^2}\gg D^2$). This means that the number
of basis states participating in a given eigenstate is large,
$N_c\sim \Gamma /D\gg 1$, or vice versa, a given basis state $k$
contributes to a large number of nearby eigenstates with energies
$|E-E_k|\sim \Gamma $. Apart from the smooth variation of
$\overline{\left| C_k^{(i)}\right| ^2}$ the statistics of the eigenstate
components $C_k^{(i)}$ is close to Gaussian \cite{Ce}.
In this situation it is appropriate to call the stationary states of the
system {\em chaotic} or {\em compound} eigenstates.

Using Eq. (\ref{LDOS}) we can re-write expression (\ref{nsE}) for the
occupation numbers in the integral form
\begin{equation}\label{nsEint}
n_s(E)=\int \overline{\left| C_k^{(i)}\right| ^2}
\langle k|\hat n_s|k\rangle \rho (E_k)dE_k \approx
\int \rho _w(E,k) \langle k|\hat n_s|k\rangle dE_k~,
\end{equation}
where we used the fact that $\rho (E_k)\approx \rho (E)$ near the maximum
of $\rho _w(E,k)$. The above representation is very similar to
Eq. (\ref{nsTint}) for $n_s(T)$.
Equation (\ref{nsEint}) describes averaging over the compound state
strength function $\rho _w(E,k)$ of width $\Gamma $ centered at energy
$E$,
whereas Eq. (\ref{nsTint}) refers to a thermal average with the $w_T(E_k)$
probability density, which peaks near $\overline{E}(T)$. Of course,
the width of the distribution $w_T(E_k)$ depends on the temperature
[$\sim aT^{3/2}$ for Eqs. (\ref{rhoE}) and (\ref{TE})], but if we choose
the temperature by setting $E=\overline{E}(T)$, the two averages $n_s(E)$
and $n_s(T)$ should be
close to each other, provided the widths of $\rho _w(E,k)$ and $w_T(E_k)$
are much greater than the multiparticle level spacing, and the difference
between these widths does not exceed the single-particle energy interval
in the system \cite{FI97}.

In the next section we calculate thermally averaged occupation numbers
and establish a relation between the effective temperature $T$ and
the excitation energy for the Ce atom. Numerical calculations will
confirm that a temperature-based statistical theory agrees with
the dynamic calculation, and describes well the peculiar behaviour of the
occupation numbers in Ce.

\section{Calculation of thermally averaged occupation numbers}
\label{sec:5}

\subsection{Statistical model}\label{stmod}

Let us now perform a statistical calculation of the occupation numbers
for a system of $n$ particles distributed over $r$ orbitals with energies
$\varepsilon _s$ and degeneracies $g_s$ ($s=1,\dots ,r$). We will assume
that the two-body interaction of any two particles in the orbitals $s$
and $p$ is $U_{sp}$, where both the direct and exchange terms are included:
\begin{equation}\label{Usp}
U_{sp}=\frac{1}{g_s(g_p-\delta _{sp})}
\sum _{\alpha \in s}\sum _{\beta \in p}
(V_{\alpha \beta \beta \alpha }-V_{\alpha \beta \alpha \beta })~.
\end{equation}
Thus, $U_{sp}$ is averaged over the degenerate single-particle states
within the orbitals $s$ and $p$.

The energy of a particular many-particle state $k$ is now given by
\begin{equation}\label{Ek}
E_k=\sum _{s=1}^{r}{\cal N}_s\varepsilon _s+\sum _{s=1}^{r}\sum _{p=s}^{r}
\frac{{\cal N}_s({\cal N}_p-\delta _{sp})}{1+\delta _{sp}}U_{sp}~,
\end{equation}
where ${\cal N}_s$ is an integer number of particles in the orbital $s$
($0\leq {\cal N}_s\leq g_s$), and $\sum _s{\cal N}_s=n$. The state $k$ is
specified by the orbital occupation numbers ${\cal N}_s$ and is
$G_k$-degenerate, where $G_k=\prod _{s=1}^{r}
\left( {g_s\atop {\cal N}_s}\right)$. Of course, Eq. (\ref{Ek})
corresponds to the ``diagonal'' approximation ($E_k\approx H_{kk}$),
since we neglect the effect of mixing of states due to the residual
interaction (off-diagonal part of the Hamiltonian). In the low-energy
part of the spectrum this interaction pushes the eigenstates down
with respect to their diagonal-approximation values because of the
usual level-repulsion effect.

In this model we cannot keep hold of the total angular momentum, so our
calculation will yield quantities averaged over various angular momenta
$J$ and projections $J_z$. However, it is relatively easy to ensure
the conservation of parity. If the orbital $s$ has a parity of
$P _s$ (either 1 or $-1$), the parity of the multiparticle state $k$ is
$\prod _{s=1}^{r}P _s^{{\cal N}_s}$. Therefore, one can easily
select multiparticle states with a given parity when calculating
statistical sums like that of Eq. (\ref{nsT}).

In the canonical ensemble the probability of finding the system in the
state $k$ is given by
\begin{eqnarray}\label{canonw}
w_k&=&Z^{-1}G_k\exp (-E_k/T)~,\\
\mbox{where}\quad Z&=&\sum _kG_k\exp (-E_k/T)~,\label{canonZ}
\end{eqnarray}
and the sum over $k$ runs over all multiparticle states, possibly with a
restriction on parity. The average occupation numbers
$n_s(T)=\overline{{\cal N}_s}/g_s$ are calculated as
\begin{equation}\label{nsT1}
n_s(T)=g_s^{-1}\sum _k{\cal N}_s^{(k)}w_k~,
\end{equation}
where ${\cal N}_s^{(k)}$ is the number of particles in the orbital $s$
in the multiparticle state $k$. In the diagonal approximation the energy
of the system is related to the temperature as $E=\overline{E}(T)$,
where
\begin{equation}\label{EavT}
\overline{E}(T)=\sum _k E_kw_k ,
\end{equation}
is the canonical average. However, one can include the off-diagonal part
of the Hamiltonian in the definition of temperature by introducing the
energy shift $\Delta _E(T)$,
\begin{equation}\label{EavT1}
E=\overline{E}(T)-\Delta _E(T) .
\end{equation}
$\Delta _E(T)$ is positive in the lower half of the spectrum, which means
that the statistical effects of interaction between particles increase
the effective temperature. At high temperatures 
$\Delta _E(T)=2<H_{ii}-E^{(i)}>$, where  $H_{ii}-E^{(i)}$ is the simple
energy shift due to the non-diagonal matrix elements of the Hamiltonian 
$H_{ik}$ . This estimate is based on the mean energy of the
components $|k\rangle $ in the eigenstate $|i\rangle $: 
$(\overline{E_k})_i \equiv \sum_k H_{kk}|C_k^{(i)}|^2=
E^{(i)}+\Delta_E $ (see \cite{FI97}).

In general, the occupation numbers obtained from Eq. (\ref{nsT1})
will be different from those predicted by the Fermi-Dirac distribution
(see Sec. \ref{numcal}).
In appendix \ref{app} we look at how the FDD
(\ref{FDd}) is derived from the canonical statistical distribution,
Eq. (\ref{nsT1}), and see what are the limitations on the
interaction between the particles for the derivation to be valid.

\subsection{Numerical calculations}\label{numcal}

To perform numerical calculations of the occupation numbers for Ce in the
statistical model outlined above we use the same
set of 14 relativistic orbitals as in the CI calculation described in
Sec. \ref{sec:3}. The orbital energies are obtained as
$\varepsilon _s=\langle s|H_c|s\rangle $, where $H_c$ is the frozen
Dirac-Fock Hamiltonian of the Ce$^{4+}$ core, and the averaged Coulomb
matrix elements $U_{sp}$ are found from Eq. (\ref{Usp}). Their numerical
values for the 7 lowest orbitals $4f_{5/2}$, $4f_{7/2}$, $6s_{1/2}$,
$5d_{3/2}$, $5d_{5/2}$, $6p_{1/2}$, and $6p_{3/2}$ are given in
Table~\ref{param}. For excitation energies below the ionization threshold
these orbitals are the most important.

Using the statistical model formulae we have calculated the occupation
numbers $n_s(T)$ (\ref{nsT1}) and the energy of the system (\ref{EavT}) as
functions of $T$ (see Figs. 3a and 3b). The relation between the
energy of the system and temperature $E=\overline{E}(T)-\Delta _E(T)$ can
be inverted and used to plot the dependence of the statistical model
occupation numbers as functions of the energy $E$. In this work we
are mostly interested in the low-energy part of the spectrum, and we
put $\Delta _E(T)=\mbox{const}$ to fit the true ground-state energy of
the system at $T=0$. In Fig. \ref{compar} we compare
the results of the statistical model with the energy-averaged occupation
numbers obtained from the CI calculation of the Ce eigenvalues and
eigenstates, Eqs. (\ref{fulldia}) and (\ref{nsE}). We see that the
complicated non-Fermi-Dirac energy dependence of the occupation numbers
in Ce is reproduced well by the statistical model.

To study the effect of the off-diagonal matrix elements of the Hamiltonian
on the temperature we calculate the canonically-averaged mean energy
of the system (\ref{EavT}) using the set of the basis-state energies
$E_k\equiv H_{kk}$ and that of the exact eigenstates $E^{(i)}$. The difference
between these two mean energies is plotted in Fig. \ref{TTT}. It is almost
constant at small temperatures and follows
\begin{equation}\label{del_e}
\Delta _E(T)\simeq \frac{\overline {\sum _{k\neq l}H_{kl}^2}}{T},
\end{equation}
at large $T$ (see \cite{FI97}).

Note that the energy shift $\Delta _E(T)$ is larger than the simple
difference between the diagonal matrix elements and exact eigenvalues
$H_{ii}-E^{(i)}$.
 This is because
the true occupation numbers  (\ref{nsE}) even in the exact ground state  are
not integer  (see Fig. \ref{compar}) due to the admixture of higher
configurations. This means that  the effective temperature of the ground state
is already not zero (see discussion in \cite{FI97}).

\section{Discussion and conclusions.}
Most importantly, the agreement observed in Fig. \ref{compar} means that
the interaction between the
particles indeed introduces some kind of equilibrium in the system
(``micro-canonical'' distribution). Moreover, averaging over it yields
results close to those over a canonical ensemble (\ref{canonw}), with the
temperature chosen to reproduce the total energy of the system. This
equivalence is always true for large system where any albeit weak
interaction between particles leads to equilibrium. However, in a
{\em few-particle} system the residual two-body interaction must be
{\em strong enough} to produce chaotic eigenstates and facilitate
statistical description (see \cite{alt,shep} where criteria for the
interaction strength are discussed).

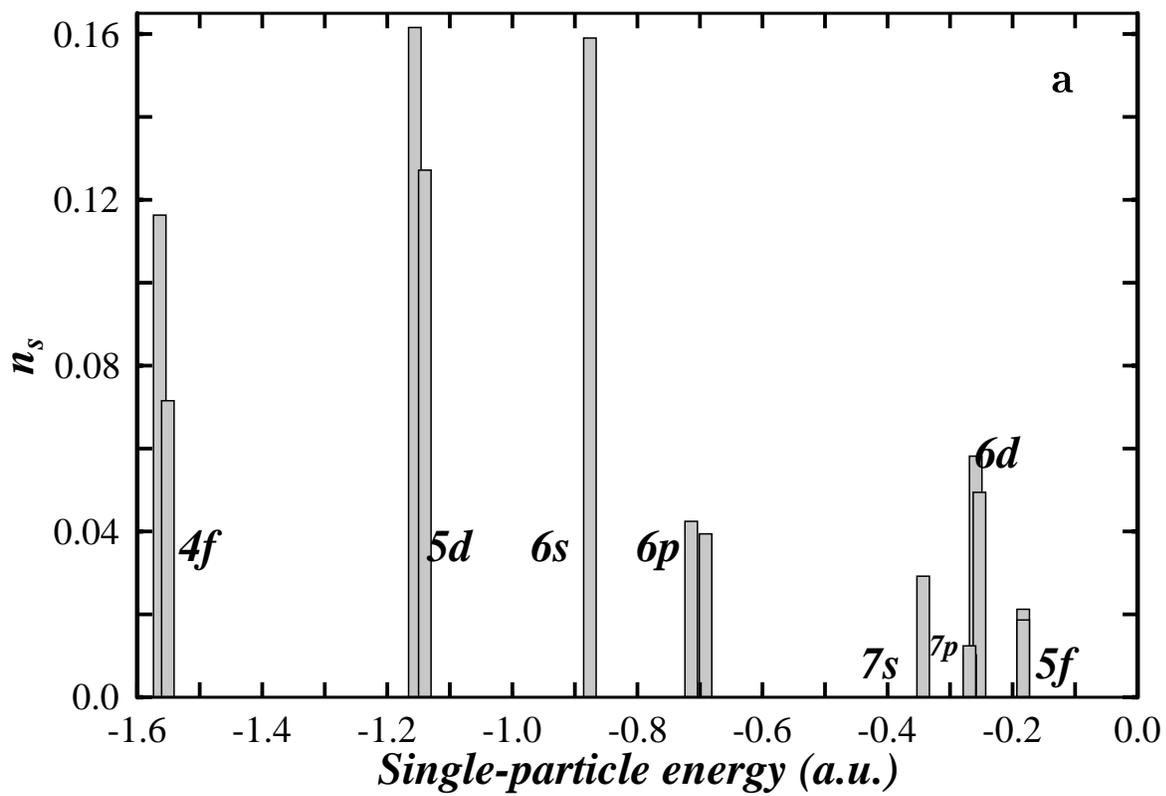
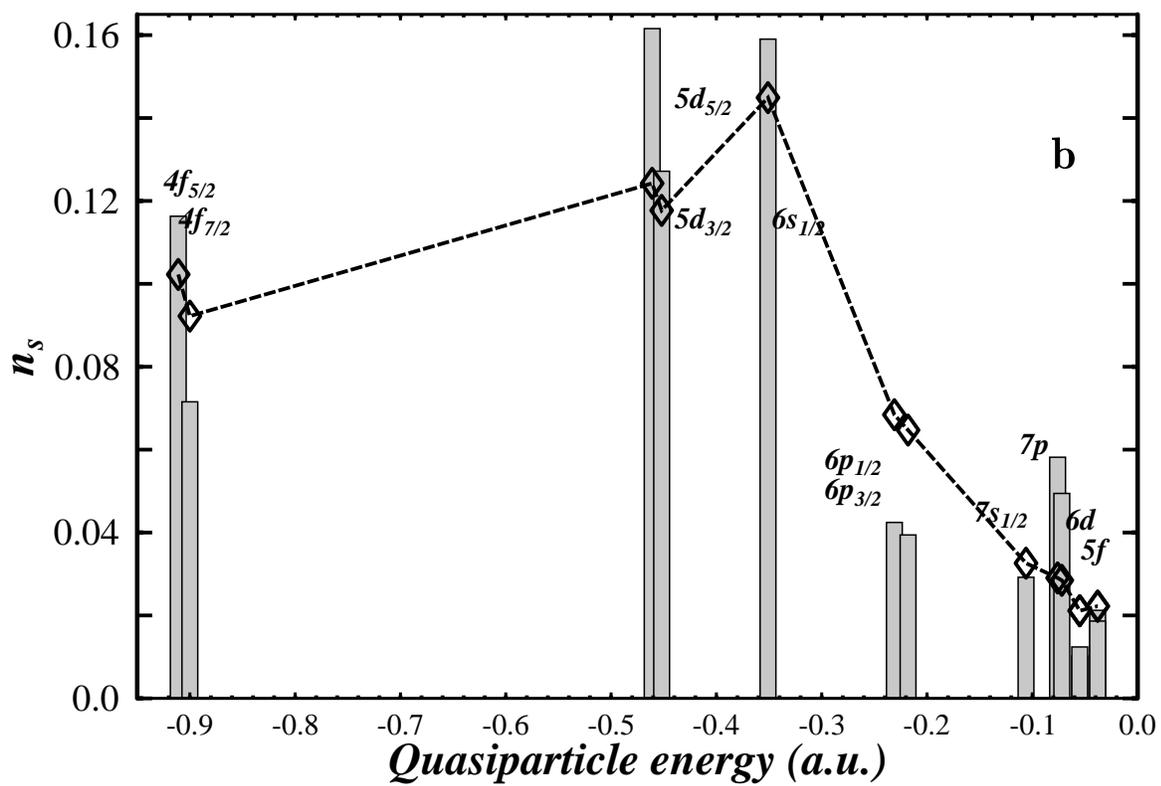
\begin{figure}
\caption{Occupation numbers $n_s(E)$ (see Eq. (\protect\ref{nsE})),
for the even states of Ce at the excitation energy of $E-E_g=4.5$ eV
versus the single-particle energies $\varepsilon _s$ of the orbitals (a),
and quasiparticle energies $\tilde \varepsilon _s$ (b). Diamonds connected
by the dashed line represent the result of our statistical calculation
with $T=0.097$ a.u. \label{nfdsp}}
\end{figure}

We would like to reiterate, that although the temperature-based description
is apparently applicable to our 4-particle system, the orbital occupancies
could not be described by the FDD (Fig. \ref{nfdsp}). The FDD is
inapplicable to our system because of the strong interaction between the
particles [second term on the right-hand side of Eq. (\ref{Ek})]. However,
the deviation from the FDD is determined not by the magnitude of $U_{sp}$,
but rather by the size of their fluctuations. To see this assume for a
moment that all two-body matrix elements are the same, $U_{sp}\equiv U$.
In this case the double sum in Eq. (\ref{Ek}) just shifts all energies by
$\frac{U}{2}N(N-1)$, and the statistical properties of the system are the
same as for noninteracting particles. If $U_{sp}$ are different for
different orbital pairs $sp$ one can still introduce some average
interaction $\overline{U}$ and subtract this ``background'' interaction
from the interaction term in Eq. (\ref{Ek}). This procedure effectively
suppresses the interaction term, since the summands in expressions like
$\sum_{s<p} (U_{sp}-\overline{U}){\cal N}_p$ have different signs. Note
that the introduction of (energy-dependent) $\overline{U}$ is equivalent
to a local mean-field approximation (see also Appendix \ref{app}). This
approximation is good if the fluctuations of $U_{sp}$ from one orbital to
another are relatively small.

An instructive example was provided in \cite{FIC96,FI97}. In these works
a model of random two-body interactions was explored numerically and a
good Fermi-Dirac-like behaviour of the occupation numbers was observed for
as little as 4 particles distributed among 11 orbitals. However, this
regime was achieved for the relatively small two-body matrix elements with
mean zero and $\mbox{r.m.s.}V\sim 0.1 d_1 $, where $d_1 $ is the mean
level spacing between the single-particle orbitals. On the other hand, for
smaller two-body interaction strengths the occupation numbers distribution
was not smooth, and did not agree well with the Fermi-Dirac formula,
because the statistical equilibrium needed was not achieved.

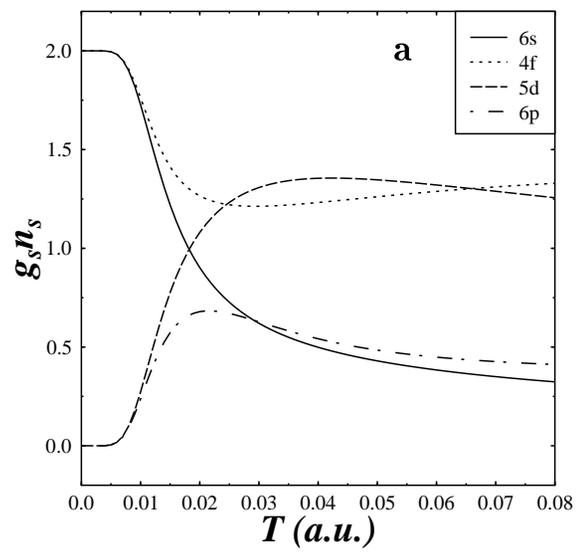
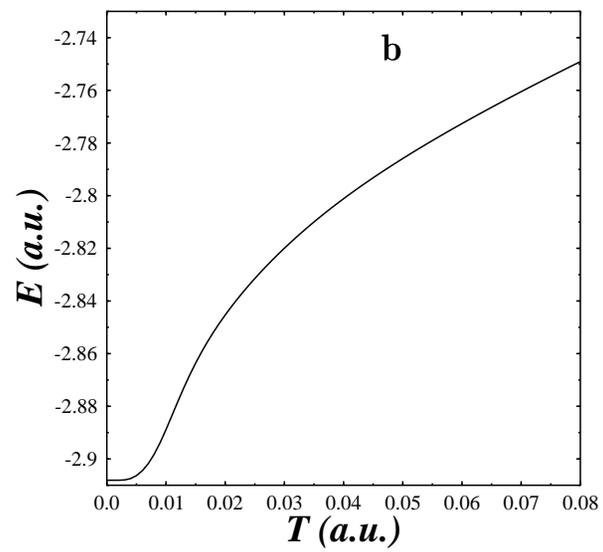
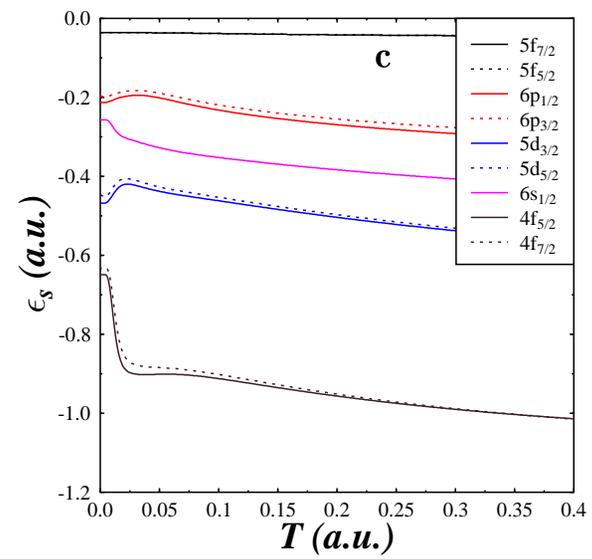
\begin{figure}
\caption{Temperature dependence of the occupation numbers for the
orbitals $g_sn_s(T)$ (a), energy of the system $\overline{E}(T)$ (b) and
quasiparticle energies $\tilde \varepsilon _s$ (c), calculated in the
statistical model of the Ce atom, Eqs.
(\protect \ref{Ek})-(\protect \ref{EavT}). The occupation numbers shown
for the $4f$, $5d$ and $6p$ subshells are sums of those of their
fine-structure sublevels $j=l\pm \frac{1}{2}$.\label{nET}}
\end{figure}

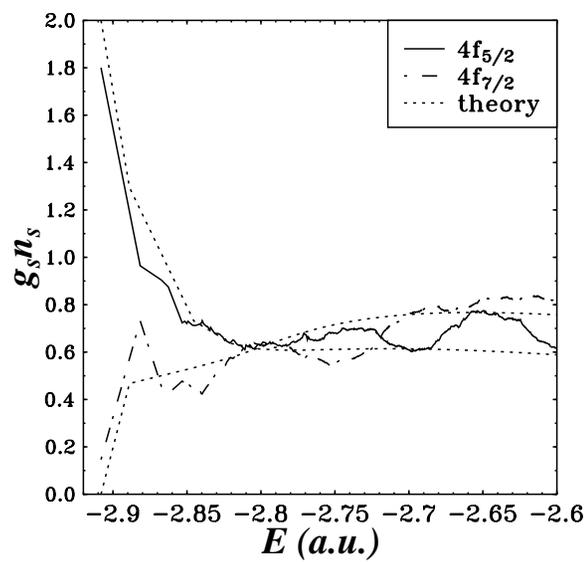
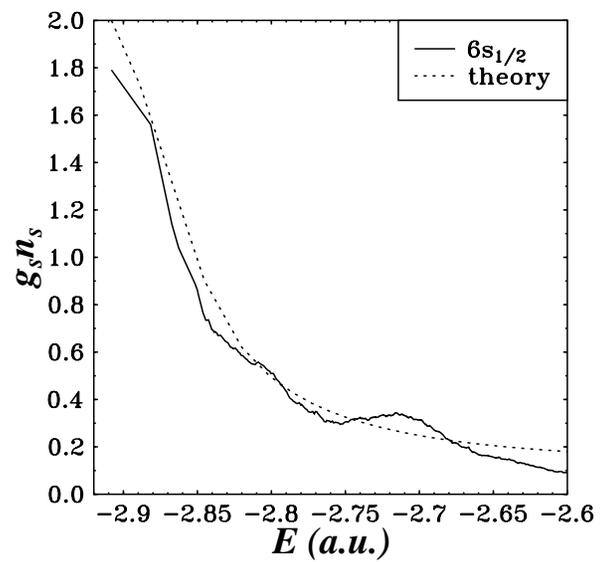
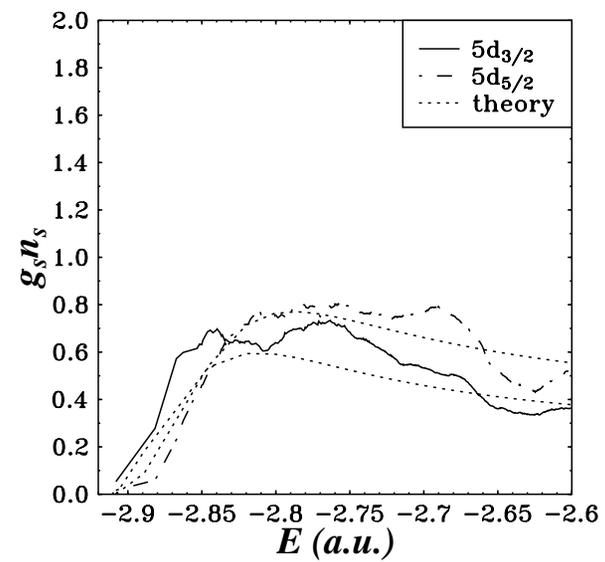
\begin{figure}
\caption{Comparison of the orbital occupancies $g_s n_s(E)$
obtained from the exact diagonalization (solid and dash-dot 
lines) with $g_sn_s(T({\overline E}))$ obtained from our statistical
theory (dotted lines).\label{compar}}
\end{figure}

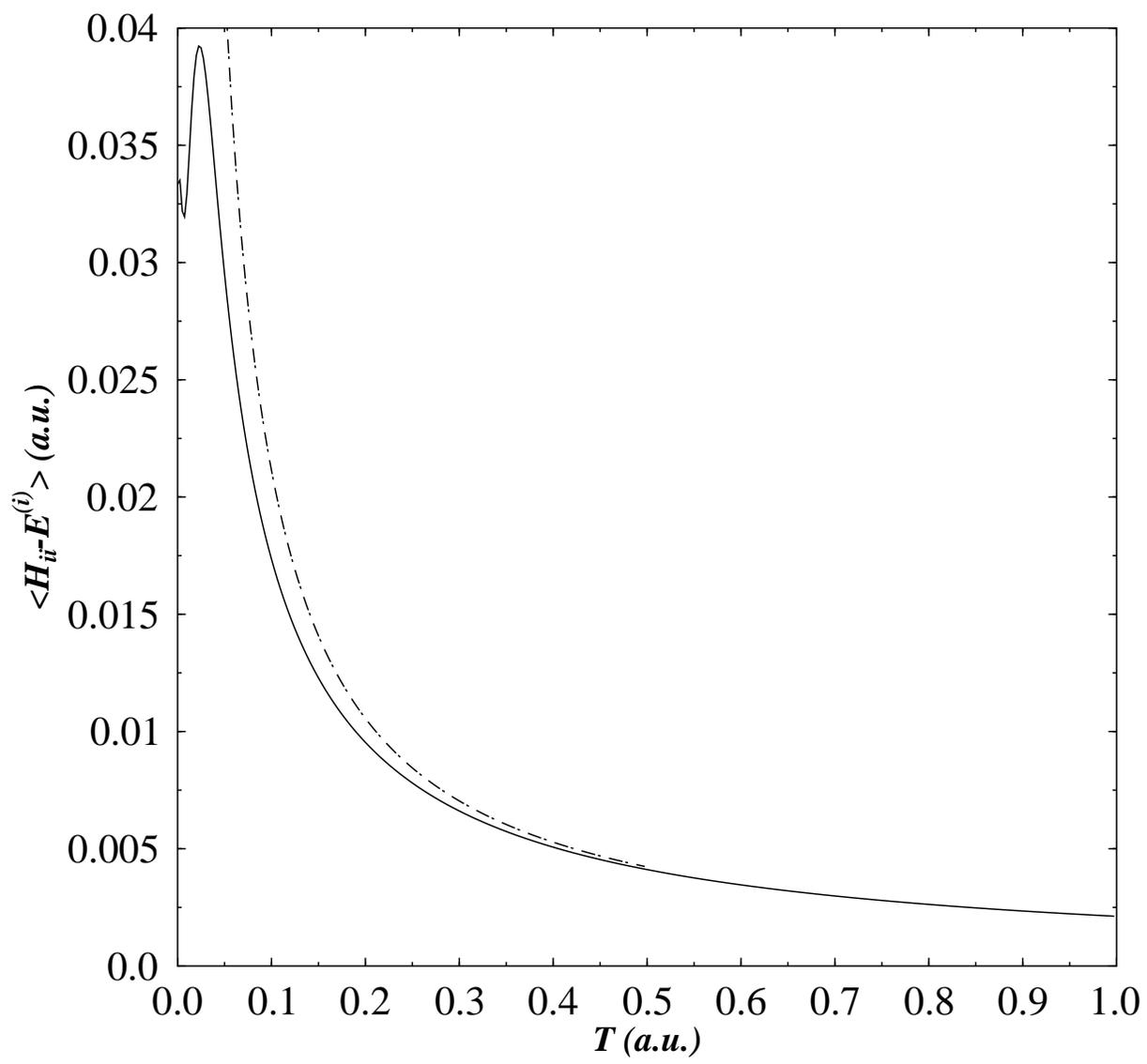
\begin{figure}
\caption{
Difference between the mean values of the energy obtained from
the canonical distribution using the energies of the basis-states,
$E_k\equiv H_{kk}$, and those of the exact eigenstates. The dash-dot line
is the high-temperature analytical approximation (\protect \ref{del_e}).
\label{TTT}}
\end{figure}

The situation in the Ce atom is different. The $4f$ orbital has a much
smaller radius than any other orbital, and the Coulomb interactions have
a hierarchy of scales:
\begin{equation}\label{hier}
U_{4f4f}> U_{4fs}> U_{sp},
\end{equation}
where $s$ and $p$ are orbitals other than $4f$.
Indeed, the Coulomb interaction between the electrons is determined
mostly by the mean radius of the largest orbital. Thus, for the two
orbitals $s$ and $p$ such that $r_s<r_p$, the Coulomb interaction
is $U_{sp}\approx e^2/r_p$ (this formula is exact for the two electrons
distributed uniformly over the surfaces of two spheres of radii $r_s$ and
$r_p$). Because of (\ref{hier}) the interaction term in Eq. (\ref{Ek})
fluctuates strongly with the change of the occupation number of the
$4f$ orbital, and in effect there is no good mean-field approximation
for the excitation spectrum of Ce. The quasiparticle orbital energies
$\tilde \varepsilon _s$ can still be obtained in the statistical model
for Ce by means of Eq. (\ref{defquasi}). They are plotted in
Fig. \ref{nET}c and show considerable variation with temperature. However,
even when we use these energies instead of the Hartree-Fock ones for
plotting the occupation numbers, the resemblance to the true FDD is only
marginally better than that in Fig. \ref{nfdsp}.

The absence of reasonably defined quasiparticle orbitals and the ensuing
distortion of the Fermi-Dirac distribution are features outside the
usual Migdal theory of normal finite Fermi systems (TFFS) \cite{Mig}.
Its breakdown in the Ce atom can be associated with the open-shell
structure of the atom (nearly degenerate ``ground state'') and a clear
lack of symmetry of the ground state with one removed particle.
As a result, single-particle excitations above the ground state do not
carry good quantum numbers (like the momentum in an
infinite Fermi system, or the angular momentum in a spherically symmetric 
finite system). Moreover, even at low energies (few eV) the
single-particle excitations have large widths associated with their decay
into multiply excited configurations (the spreading width
$\Gamma \sim 2$ eV \cite{Ce}), largely because such decay is not really
limited by any selection rules (only the trivial total angular momentum
and parity are conserved).

It has been proposed in \cite{FI97} that for finite Fermi systems similar
to the Ce atom, characterized by the dense spectra of chaotic multiparticle
eigenstates, a statistical theory alternative to the standard TFFS
can be developed based on the properties of these eigenstates. Most
importantly, the existence of the chaotic eigenstates and the equilibrium
this introduces in the system is ensured by the sufficiently strong
interaction between particles.

This concept of the interaction-driven equilibrium is supported by our
present results. We have shown that this equilibrium can be described in
terms of usual statistical parameters, such as the temperature, even though
some of the system's properties are very different from those usually
expected in Fermi systems. For example, the statistics of the occupation
numbers cannot be described by the Fermi-Dirac formula.

\acknowledgments
We would like to thank M. Kozlov and V. Zelevinsky for useful discussions,
and acknowledge support of the present work by the Australian Research
Council.

\appendix

\section{ Distribution of occupation numbers for a canonical ensemble of
finite systems of interacting fermions}\label{app}

Consider a quantum system which consists of a number of single-particle
discrete states $\alpha $  (we will also call them orbitals here) with
energies $\varepsilon _\alpha $ ($\alpha =1,\dots ,m$) filled with
$n<m$ Fermi particles. The multiparticle
states $k$ of the system are identified by specifying the occupation
numbers $n_\alpha =0$ or 1 of the orbital $\alpha $,
$\sum _\alpha n_\alpha =n$. The total number of multiparticle states
$N=\left( {m\atop n}\right) $ is quite large, even for moderate
$m$ and $n$. If we allow for the interaction between the
particles, the energy of the state $k$ is given by
\begin{equation}\label{apEk}
E_k=\sum _\alpha \varepsilon _\alpha n_\alpha +\frac{1}{2}
\sum _{\alpha \beta }U_{\alpha \beta }n_\alpha n_\beta ~,
\end{equation}
where $U_{\alpha \beta }$ includes both the direct and exchange interaction
between the particles in $\alpha $ and $\beta $, and
$U_{\alpha \alpha }=0$.

Of course, the states $k$ are not eigenstates of the system. However,
if the interaction between the particles is not too strong, we can use them
for averaging over the canonical ensemble with probabilities
$w_k=Z^{-1}\exp (-E_k/T)$, where $Z=\sum _k \exp (-E_k/T)$
is the partition function, and the sum runs over all $N$ multiparticle
states \cite{note}. It will be convenient to show explicitly that
the partition function depends on the energies of the orbitals
$\varepsilon _1,\dots ,\varepsilon _m\equiv \{\varepsilon \}$, number of
particles $n$, and temperature $T$: $Z\equiv Z(\{\varepsilon \},n,T)$ (and,
strictly speaking, on the interactions $U_{\alpha \beta }$, if they are not
zero).

Using the canonical probabilities we can express the mean occupancy of the
orbital $\alpha $ as
\begin{eqnarray}\label{apnal1}
\overline n _\alpha &=&\sum _kn_\alpha ^{(k)}w_k=
\frac{\sum _kn_\alpha ^{(k)}\exp (-E_k/T)}{\sum _k \exp (-E_k/T)}\\
\label{apnal2}
&=&\frac{\sum _{k(\alpha )}\exp (-E_k/T)}
{\sum _{k(\alpha )}\exp (-E_k/T)+\sum _{k(\bar \alpha )}
\exp (-E_k/T)}~,
\end{eqnarray}
where $n_\alpha ^{(k)}=1$ or 0, depending on whether $\alpha $ is occupied
or empty in the state $k$, and we split the sum over $k$ into two sums
over the states $k(\alpha )$ where $\alpha $ is occupied, and
$k(\bar \alpha )$, where it is empty.

Let us first consider the case of
noninteracting particles ($U_{\alpha \beta }=0$). It is easy to see that
the first sum is then equal to $Z(\{\varepsilon \}' _\alpha ,n-1,T)
\exp (-\varepsilon _\alpha /T)$ and the second one is
$Z(\{ \varepsilon \}' _\alpha ,n,T)$, where
$\{\varepsilon \}' _\alpha $ is the set of $m-1$ orbitals obtained by
discarding $\alpha $ from $\{\varepsilon \}$. Equation can then be written
as
\begin{equation}\label{FDZ}
\overline n _\alpha =\left[ 1+
\frac{Z(\{ \varepsilon \}' _\alpha ,n,T)}
{Z(\{\varepsilon \} _\alpha ,n-1,T)}
\exp (\varepsilon _\alpha /T)\right] ^{-1} ~.
\end{equation}
This equation is very similar to the Fermi-Dirac formula (\ref{FDd}),
if we introduce the chemical potential $\mu $ by
\begin{equation}\label{ratZ}
\frac{Z(\{ \varepsilon \}' _\alpha ,n,T)}
{Z(\{\varepsilon \}' _\alpha ,n-1,T)}\equiv \exp (-\mu /T)~.
\end{equation}
The problem is that this ratio on the left-hand side in fact depends on
which orbital $\alpha $ is deleted from the set
$\{ \varepsilon \}$ to form $\{ \varepsilon \}' _\alpha $, and so does the
``chemical potential'' $\mu $. If we write
Eq. (\ref{FDZ}) for $\overline n_\beta $ with $\beta \neq \alpha $, the
set $\{\varepsilon \}' _\beta $ will produce a different ratio
$Z(\{ \varepsilon \}' _\beta ,n,T)/Z(\{\varepsilon \}' _\beta ,n-1,T)$,
and as a result, a different value of $\mu $.
However, $\{\varepsilon \}' _\beta $
can be obtained from $\{\varepsilon \}' _\alpha $ by simply moving the
orbital energy from $\varepsilon _\beta $ to $\varepsilon _\alpha $.
So, the difference between the values of $\mu $ for different orbitals
can be probed by calculating the derivative of Eq. (\ref{ratZ}) with
respect to the energy of some orbital $\beta $. Using the relation
$\overline n _\beta =-T\partial Z/\partial \varepsilon _\beta $, valid
for noninteracting particles (see Eqs. (\ref{apEk}) and (\ref{apnal1}))
we obtain
\begin{equation}\label{dermu}
\frac{\partial \mu }{\partial \varepsilon _\beta }=
\overline n_\beta ^{(n)}-\overline n_\beta ^{(n-1)} ~,
\end{equation}
where $\overline n_\beta ^{(n)}$ and $\overline n_\beta ^{(n-1)}$ are
the mean occupation numbers for $n$ and $n-1$ particles distributed among
the $\{\varepsilon \}' _\alpha $ orbital set. It is obvious that the
right-hand side of Eq. (\ref{dermu}) is larger near the Fermi level,
$|\varepsilon _\beta -\mu |\lesssim T$, and is almost zero outside this
interval. We can estimate that the total difference between the
value of $\mu $ for orbitals well below the Fermi level and those well
above it is
\begin{equation}\label{difmu}
\delta \mu =\int \left( n_\beta ^{(n)}-\overline n_\beta ^{(n-1)}
\right) d\varepsilon _\beta =d_1
\int \left( n_\beta ^{(n)}-\overline n_\beta ^{(n-1)}\right)
\frac{d\varepsilon _\beta }{d_1}\approx d_1
[n-(n-1)]=d_1 ~,
\end{equation}
where $d_1$ is the mean spacing between the single-particle
orbitals. Thus, in a system with discrete orbital energies the chemical
potential can be considered as constant to within
$\sim d_1$ accuracy. At finite temperatures the width of
the smoothened Fermi-Dirac ``step'' is of the order of $T$, therefore
$\mu =\mbox{const}$ is valid for $T\gg d_1 $ (or for $\mu \gg d_1$).
 Note that this condition means that the number of orbitals within the 
Fermi-Dirac ``step'' is large.

For the interacting particles the sum
$\sum _{k(\alpha )}\exp (-E_k/T)$ which gives rise to
$\exp( \varepsilon _\alpha /T)$ in Eq. (\ref{FDZ}) can be written as
\begin{eqnarray}
\sum _{k(\alpha )}\exp (-E_k/T)&=&\sum _{k(\alpha )}\exp \left[
-\frac{1}{T}\left( \sum _{\beta \neq \alpha }\varepsilon _\beta n_\beta
+\frac{1}{2} \sum _{\beta , \gamma \neq \alpha }U_{\beta \gamma }
n_\beta n_\gamma \right)\right] \nonumber \\ \label{expE}
&\times &\exp \left[ -\frac{1}{T}\left( \varepsilon _\alpha +
\sum _{\beta \neq \alpha }U_{\alpha \beta }n_\beta \right)\right] ~.
\end{eqnarray}
In this formula the last exponent contains the energy of the particle
in $\alpha $, which depends on the occupancies $n _\beta $ of the other
orbitals. When summation over $k(\alpha )$ is carried out the exponent
is averaged over different distributions of $n-1$ particles among all
orbitals but $\alpha $. The result can be presented approximately as
\begin{equation}\label{deftileps}
Z(\{\varepsilon \} _\alpha ,n-1,T)\exp (-\tilde \varepsilon _\alpha /T) ~,
\end{equation}
where we replaced the averaged exponent by the exponent containing the
{\em quasiparticle} energy,
\begin{equation}\label{defquasi}
\tilde \varepsilon _\alpha =\varepsilon _\alpha +
\sum _{\beta \neq \alpha }U_{\alpha \beta }\overline n_\beta ~,
\end{equation}
and the mean values $\overline n_\beta $ are, strictly speaking, different
from those from Eq. (\ref{apnal1}), as one particle has always been kept
in $\alpha $ in the sum (\ref{expE}). Therefore, the ``Fermi-Dirac''
anzats (\ref{FDZ}) holds for the interacting particles, if we replace
the single-particle energies $\varepsilon _\alpha $ with the
temperature-dependent quasiparticle energies
$\tilde \varepsilon _\alpha $ from Eq. (\ref{defquasi})

Note that the transformation of
Eq. (\ref{expE}) into Eqs. (\ref{deftileps}), (\ref{defquasi}) is exact
up to first order in $\overline n_\beta $, and to all orders, if
$U_{\alpha \beta }\equiv U$ for all orbitals. In the latter case
$\tilde \varepsilon _\alpha =\varepsilon _\alpha +(n-1)U$ is a trivial
redefinition of the single-particle energies. That is why replacing
$\varepsilon _\alpha $ with the quasiparticle energies
$\tilde \varepsilon _\alpha $ is a valid operation, unless
the interactions $U_{\alpha \beta }$ fluctuate strongly, and the number
of active particles is small. In the latter case the mean-field
approximation is inadequate and the introduction of quasiparticles is
not very meaningful.

\begin{table}
\caption{Single-particle energies and Coulomb matrix elements for the
valence and lowest excited orbitals in Ce.}
\label{param}
\begin{tabular}{ccccccccc}
Orbital & $\varepsilon _s$ & \multicolumn{7}{c}{Coulomb matrix elements
$U_{sp}$, a.u.} \\
\cline{3-9}
$nlj$ & a.u. & $4f_{5/2}$ & $4f_{7/2}$ & $6s_{1/2}$ & $4d_{3/2}$ &
$4d_{5/2}$ & $6p_{1/2}$ & $6p_{3/2}$ \\
\hline
$4f_{5/2}$ & $-$1.564 & 0.791 & 0.800 & 0.260 & 0.423 &0.422&0.223&0.216 \\
$4f_{7/2}$ & $-$1.551 & 0.800 & 0.787 & 0.259 & 0.428 &0.416&0.223&0.215 \\
$6s_{1/2}$ & $-$0.876 & 0.260 & 0.259 & 0.199 & 0.231 &0.230&0.162&0.158 \\
$4d_{3/2}$ & $-$1.156 & 0.423 & 0.428 & 0.231 & 0.330 &0.331&0.200&0.198 \\
$4d_{5/2}$ & $-$1.141 & 0.422 & 0.416 & 0.230 & 0.331 &0.325&0.204&0.195 \\
$6p_{1/2}$ & $-$0.714 & 0.223 & 0.223 & 0.162 & 0.200 &0.204&0.169&0.157 \\
$6p_{3/2}$ & $-$0.691 & 0.216 & 0.215 & 0.158 & 0.198 &0.195&0.157&0.156 \\
\end{tabular}
\end{table}

\begin{thebibliography}{99}

\bibitem{Ce} V. V.~Flambaum, A. A. Gribakina, G. F. Gribakin, and
M. G. Kozlov, Phys. Rev. A. {\bf 50}, 267 (1994); A. A. Gribakina,
V. V. Flambaum, and G. F. Gribakin, Phys. Rev. E. {\bf 52}, 5667 (1995).
\bibitem{Zel}M. Horoi, V. Zelevinsky, and B. A. Brown, Phys. Rev. Lett.
{\bf 74}, 5194 (1995); Phys. Lett. B {\bf 350}, 141 (1995);
V.~Zelevinsky, B. A.~Brown, N.~Frazier, and M.~Horoi, Phys. Rep.
{\bf 276}, 85 (1996).
\bibitem{FIC96} V. V. Flambaum, F. M. Izrailev, and G. Casati,
Phys. Rev. {\bf 54}, 2136 (1996).
\bibitem{FI97} V. V.~Flambaum, F. M.~Izrailev, Phys. Rev. E {\bf 55}, R13
(1997); V. V.~Flambaum, F. M.~Izrailev, Phys. Rev E {\bf 56 }, 5144 (1997).
\bibitem{PiNo}D. Pines and P. Nozi\'eres, {\em The Theory of Fermi Liquids}
(Benjamin, New York, 1966).
\bibitem{LL5}L. D. Landau and E. M. Lifshitz, {\em Statistical physics}
(Pergamon Press, New York, 1969).
\bibitem{tables} W. C. Martin, R. Zabulas and L. Hagan, {\em Atomic
Energy Levels - The Rare-Earth Elements}, Natl. Bur. Stand. Ref. Data
Ser., Natl. Bur. Stand. (U.S.), NBS-60, (U.S., GPO, Washington, DC,
1978).
\bibitem{AIS} V. V. Flambaum, A. A. Gribakina, and G. F. Gribakin,
Phys. Rev. A {\bf 54}, 2066 (1996).
\bibitem{corex} Of course, if one needs good qualitative results for the
spectrum of Ce, virtual core excitations must be taken into account. This
can be done, for instance, by introducing the effective correlational
electron-core potential and the screened Coulomb interaction between the
valence electrons using many-body theory methods, V. A. Dzuba,
V. V. Flambaum, and M. G. Kozlov, Phys. Rev. A, {\bf 54}, 3948 (1996).
\bibitem{BM} A. Bohr, and B. Mottelson, {\em Nuclear Structure}
(Benjamin, New York, 1969), Vol. 1.
\bibitem{French}It is known that the level density in the $n$-particle
system with $k$-body interaction becomes Gaussian for $n\gg k$, see
J. B. French and S. S. M. Wong, Phys. Lett. B {\bf 35}, 5 (1970);
O. Bohigas and J. Flores, Phys. Lett. B. {\bf 34}, 261 (1971).
\bibitem{FV93}This analogy was clearly stated by V. V. Flambaum and
O. K. Vorov, Phys. Rev. Lett. {\bf 70}, 4051 (1993).
\bibitem{wigner}E. P. Wigner, Ann. Math. {\bf 62}, 548 (1955); {\em ibid}
{\bf 65}, 203 (1957).
\bibitem{CFIC}Y. V. Fyodorov, O. A. Chubykalo, F. M. Izrailev, and G.
Casati, Phys. Rev. Lett. {\bf 76}, 1603 (1996).
\bibitem{alt}B. L. Altshuller, Y. Gefen, A. Kamenev, and L. S. Levitov,
Phys. Rev. Lett. {\bf 78}, 2803 (1997).
\bibitem{shep} P. Jacquod and D. L. Shepelyansky, Phys. Rev. Lett.
{\bf 79}, 1837 (1997); B. Georgeot and D. L. Shepelyansky, Preprint
cond-mat/9707231 (1997).
\bibitem{Mig}A. B. Migdal, {\em Theory of Finite Fermi-Systems,
and Applications to Atomic Nuclei} (New York: Interscience Pub.
1967).
\bibitem{note}The eigenstates of the system are in fact linear combinations
of the states $k$, which can be considered as multiparticle basis states
of the system in the Fock space. The mixing of these basis states within
the eigenstates occurs {\em locally}, in the energy interval determined
by the spreading width $\Gamma $, which depends on the strength of the
residual interaction between the particles and practically does not
depend on energy. It is usually narrower than the typical interval
involved in averaging over the canonical ensemble, which increases as
$T^{3/2}$ [this estimate follows from Eq. (\ref{rhoE})].



\bibitem{exact}
In numerical calculation we have always to cut-off Hamiltonian matrix 
somewhere. Therefore, we intend by the exact states here the
eigenstates of the diagonalized matrix in the truncated basis. 
\bibitem{deg}
Degeneracy $g_s$ of $4f$ and $5d$ levels are 14 and 10 correspondingly but
the number of active particle  $N=4\ll g_s$ in our case.
\end{thebibliography}
\end{document}